\documentclass[aps,prb,preprint]{revtex4}
\usepackage{graphicx}
\usepackage{xcolor}
\begin{document}
\title{Multifunctional acoustic device based on phononic crystal with independently controlled asymmetric rotating rods}
\author{Hyeonu Heo}
\altaffiliation{Current address: Pennsylvania State University, Graduate Program in Acoustics, University Park, PA 16802.}
\affiliation{Department of Physics, University of North Texas, Denton, TX 76203, USA}
\author{Arup Neogi}
\email{arup@uestc.edu.cn}
\affiliation{Institute of Fundamental and Frontier Sciences, University of Electronic Science and Technology of China, Chengdu 611731, China}
\author{Zhiming Cui}
\author{Zhihao Yuan}
\author{Yihe Hua}
\author{Jaehyung Ju}
\email{jaehyung.ju@sjtu.edu.cn}
\affiliation{UM-SJTU Joint Institute, Shanghai Jiao Tong University, Shanghai 200240, China}
\author{Ezekiel Walker}
\affiliation{Echonovus Inc., Denton, TX 76205, USA}
\author{Arkadii Krokhin}
\email{arkady@unt.edu}
\affiliation{Department of Physics, University of North Texas, Denton, TX 76203, USA}

\date{\today }


\begin{abstract}
A reconfigurable  phononic crystal (PnC) is proposed where elastic properties can be modulated by rotation of asymmetric solid scatterers immersed in water. The scatterers are metallic rods with cross-section of $120^{\circ}$ circular sector. Orientation of each rod is independently controlled by an external electric motor that allows continuous variation of the local scattering parameters and dispersion of sound in the entire crystal.
Due to asymmetry of the scatterers, the crystal band structure  possesses highly anisotropic bandgaps. Synchronous rotation of all the scatterers by a definite angle changes regime of reflection to regime of transmission and vice versa. The same mechanically tunable structure functions as a gradient index medium by incremental, angular reorientation of rods along both row and column, and, subsequently, can serve as a tunable acoustic lens, an acoustic beam splitter, and finally an acoustic beam steerer.

\end{abstract}

\maketitle

\section{Introduction}

The phononic crystal (PnC) is an effective tool to manipulate acoustic and elastic waves. This tool becomes more powerful if the parameters of the PnC can be tuned or tailored as needed. Different designs have been proposed to realize reconfigurable PnCs, i.e., a periodic medium where elastic properties can be changed. While elasticity is a relatively stable characteristic of a material,  it can be tuned (within a relatively narrow region) by changing temperature of the environment  or applying external fields. Ferroelectric materials exhibit variation of elastic constants at the phase transition point at the Curie temperature, leading to thermally tunable phononic band structures \cite{Jim}. Poly (N-isopropyl acrylamide) (PNIPAm) hydrogels belong to a class of smart materials where speed of sound is changed by $\sim 20\%$ within a narrow range of few degrees, where the hydrogels undergoes a volumetric phase transition \cite{Jin1}. This variation of the elastic parameters of hydrogels has been explored in fabrication of acoustic lens with tuneable focal length \cite{Walker1}. Temperature variation of the hydrogel with immersed  rods can be executed by illumination from an external infrared source \cite{Walker2} or radio \cite{Walker3} waves.

\medskip

Application of an external electric/magnetic field changes elastic properties of piezoelectric/piezomagnetic materials. It turns out that  magnetostrictive scatterers fabricated from Terfenol-D allow tuning of elastic moduli by $\sim 50\%$ under application of a magnetic field up to 2T. Details of the band structure and evolution of the bandgaps with magnetic field strength in a magnetoacoustic PnC were studied in Ref. [\onlinecite{Rob}]. Recently, a useful application of reconfigurable phononic crystal actuated by piezoelectric strain has been demonstrated for the purpose of operation with quantum memory \cite{Taylor}. Active and electrically reconfigurable metamaterial slab of deeply subwavelength thickness was developed and fabricated in Ref. [\onlinecite{Popa}]. The slab contains a chain of periodically arranged unit cells of piezoelectric membranes. Elastic properties of each membrane are controlled by electrical circuit. The slab operates as a thin lens and beam steering device. An obvious advantage of this electrically configurable system is relatively fast ($\sim$ several milliseconds) time of reconfiguration.
\begin{figure}
\includegraphics[width=\linewidth]{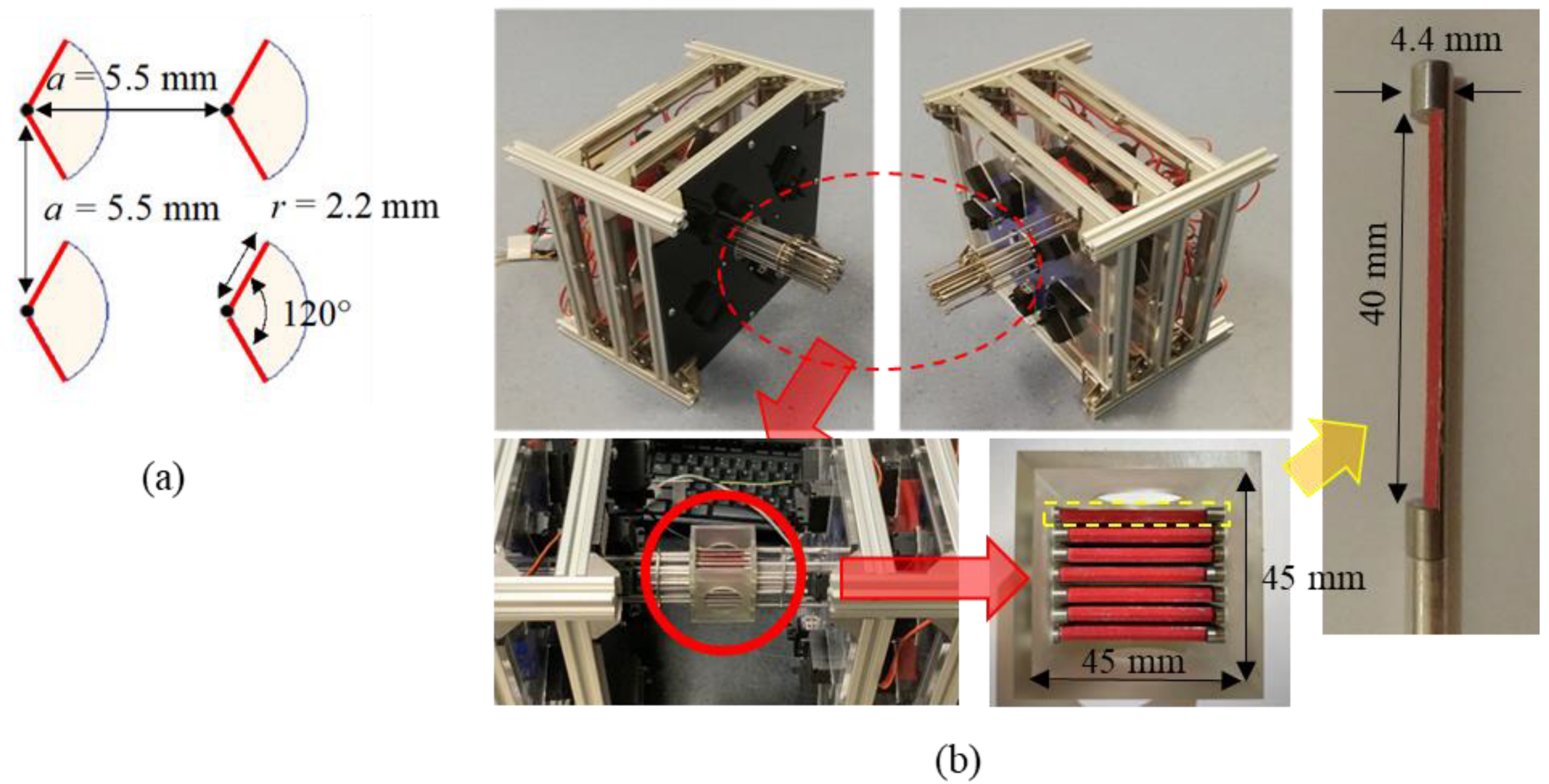}
\caption{Mechanically tunable phononic crystal device. The stainless-steel scatterers are connected to servomotors mounted to two carriers. (a) Cross-section of asymmetric PnC scatterers and geometrical parameters of the unit cell. (b) Two equal parts (bifurcated carriers) forming the device (top) and the assembled structure with the open-top 3D printed container (bottom). Red lines in (a) and red rods in (b) are the sandpaper bands attached to the rectangular parts of each rod to obtain rough surface with mean square height ~140 µm. }
\label{fig1}
\end{figure}

Mechanical variation of a PnC's unit cell may lead to essential modification of  band structure features since variation is usually accompanied by symmetry transformation.The first 3D phononic crystal with time-dependent parameters was experimentally realized in 2014 using colloidal structures. In these crystals, naturally self-assembled particles were shifted by acoustic radiation forces \cite{Caleap}.

For this work, our interests lie in reconfigurable PnCs where geometry can be continuously changed by rotation of solid rods. The band structure of a 2D PnC with rotating rods arranged in a square lattice and having square cross-section was numerically calculated and variation of the width of the bandgap with angle of rotation demonstrated \cite{Gof}.  Modulation of sound refraction from positive to negative values, depending on the angle of incidence, was experimentally realized in Ref. [\onlinecite{Feng}]. For both works, the rods were immobile, but the sample rotated as a whole with respect to the direction of the incoming wave.  Subsequently, a variety of acoustic effects were predicted and experimentally verified including multi-refringence, beam collimation, angle-dependent index of refraction, and anomalous dispersion caused by non-equivalency of differing crystallographic directions in asymmetric periodic elastic structures \cite{Buc}. Alternative classes of mechanically reconfigurable structures based on combinations of solid gears connected by solid links has also been proposed and experimentally realized \cite{Vitelli}. Due to the rotational degree of freedom of each gear, these mechanically robust structures allow smooth tuning of their mechanical characteristics over a wide range and can even produce mechanical topological insulators. Reviews of numerous reconfigurable PnCs can be found in Ref. [\onlinecite{Deymier}].

Change of temperature or global application of either external fields or pressure affects the mechanical properties of the PnC background while the geometry of the unit cell remains unchanged. In this approach, tunability is restricted by the value of acoustic contrast between the scatterers and background since external perturbations conserve the symmetry of the Bloch waves. Tunability based on modifications of the unit cell may have much wider range since it affects the structure and symmetry of the interference pattern.
Switching between different crystal lattices can be achieved by applying external load to unstable porous structures \cite{Shan} or utilizing folding transformations in origami structures \cite{Thota}. This approach allows discrete tuning of phononic crystal. Here we propose a mechanism of continuous tuning based on mechanical rotation of scatterer inside unchanged unit cell. We explore the impact of independent rotational freedom of each asymmetric scatterer while demonstrating multi-functionality of a 2D phononic crystal. The rods are arranged in a square lattice with a circular $120^{\circ}$ sector cross-section for each rod. Due to asymmetry of the rods, the band structure of the PnC can be smoothly altered when they synchronically rotate, keeping the same orientation. This design results in a metamaterial with tunable and anisotropic index of refraction that allows steering of an incident sound beam. If the orientation gradually changes from row to row, the propagating wave undergoes continuous refraction, thus realizing a gradient index medium which can be used for focusing of sound. Finally, if the rods are randomly oriented along the direction of sound propagation, random scattering causes Anderson localization of the incident wave \cite{Dhillon}.

Active mechanical tuning requires an external source of energy and usually allows variation of the effective parameters of PnC within a relatively wide range, for example index of refraction, acoustic impedance, etc.  Passive tuning schemes may be realized in nonlinear elastic materials where the propagating wave by itself changes the effective parameters depending on the wave amplitude. Nonlinear tuning does not require an external source of energy, but may require precompression to generate the necessary level of nonlinearity in granular materials. Nonlinear wave evolution may lead to spatial concentration of acoustic energy, i.e., focusing and formation of sound bullets \cite{Deng}. Passive schemes usually allow more narrow ranges of tunability than active ones. At the same time, various mechanisms of nonlinearity existing in elastic media give rise to numerous interesting effects predicted and observed in metamaterials exhibiting nonlinear elasticity \cite{Matlack}.

\section{Experimental setup}

\begin{figure}
\includegraphics[width=\linewidth]{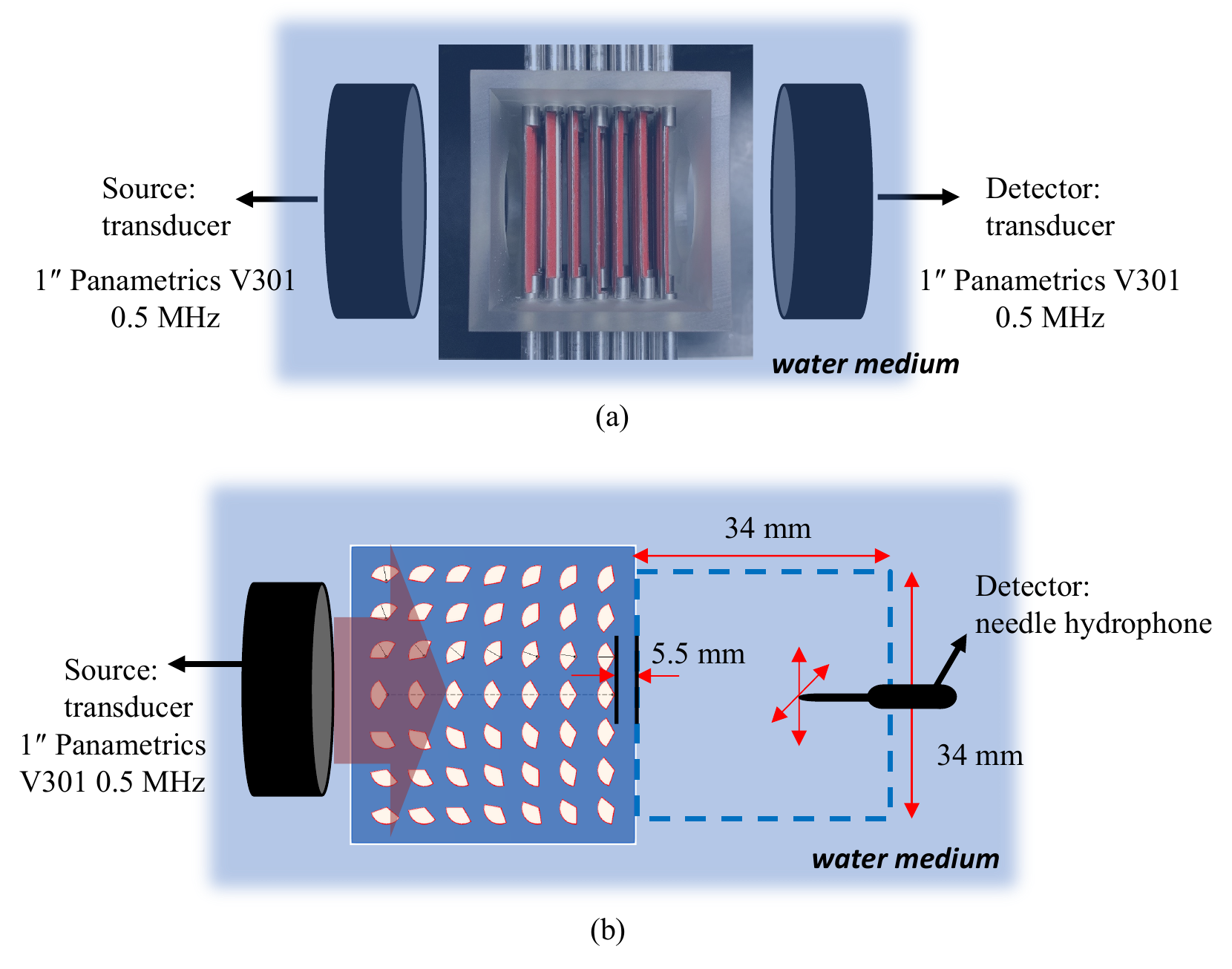}
\caption{Underwater experiment setup for (a) bistatic measurements of the transmission spectrum and (b) mapping of sound intensity using a needle hydrophone. }
\label{fig2}
\end{figure}
The reconfigurable 2D phononic crystal is shown in Figure \ref{fig1}. It is represented by a mechanically tunable $7\times7$ array of stainless steel rods.  Each rod is a sector of a circle with radius of 2.2 mm and central angle of $120^{\circ}$. Both flat surfaces of each rod are covered by sandpaper. This increases viscous dissipation within the narrow boubdary layer formed around each scatterer, thus enhancing dissipation-induced nonreciprocity. Sandpaper bands in Fig. \ref{fig1} are highlighted in red. The rods form a square lattice of period 5.5 mm, which is comparable to the wavelength of sound within the 300-550 kHz frequency range. Each rod is connected to a servo motor via micro-gear systems. The servo motor's size and physical limitations dictate a bifurcated design, where each half of the structure contains a set of the scatterer rods, motors, and control electronics. The bifurcated structure is assembled with a 3D printed container for application in a water environment.

\medskip

Transmission measurements were performed for a planar, ultrasonic beam of normal incidence on the PnC. The experimental setup is shown in Figure \ref{fig2}.  In bistatic configuration, transmission spectrum was obtained using two Olympus $1^{\prime \prime}$  Panametric V301 transducers.  Spatial distribution of intensity in the transmitted wave is measured using a needle hydrophone, see Figure \ref{fig2}b. The experimental results are compared to the corresponding numerical analysis.

\section{Results}
\subsection{Acoustic Band Structure}

COMSOL Multiphysics software was used to calculate the dispersion relation, isofrequency contours, transmission spectra of the phononic crystal, and intensity maps within the frequency range of 300 to 550 kHz. The linearized Navier-Stokes and solid mechanics modules were used to model the water background and metal scatterers, respectively. The periodic boundary conditions were applied in calculations of the dispersion relations and equifrequency contours. The transmission spectra and intensity maps were calculated for the dimensions and arrangement of those in the experiment. The transducers used in the experiment were replicated by 1 inch straight segments on opposite sides of the structure. Perfectly matched layers were implemented at the boundaries of the entire domain to avoid backscattering.

\begin{figure}
\includegraphics[width=16 cm]{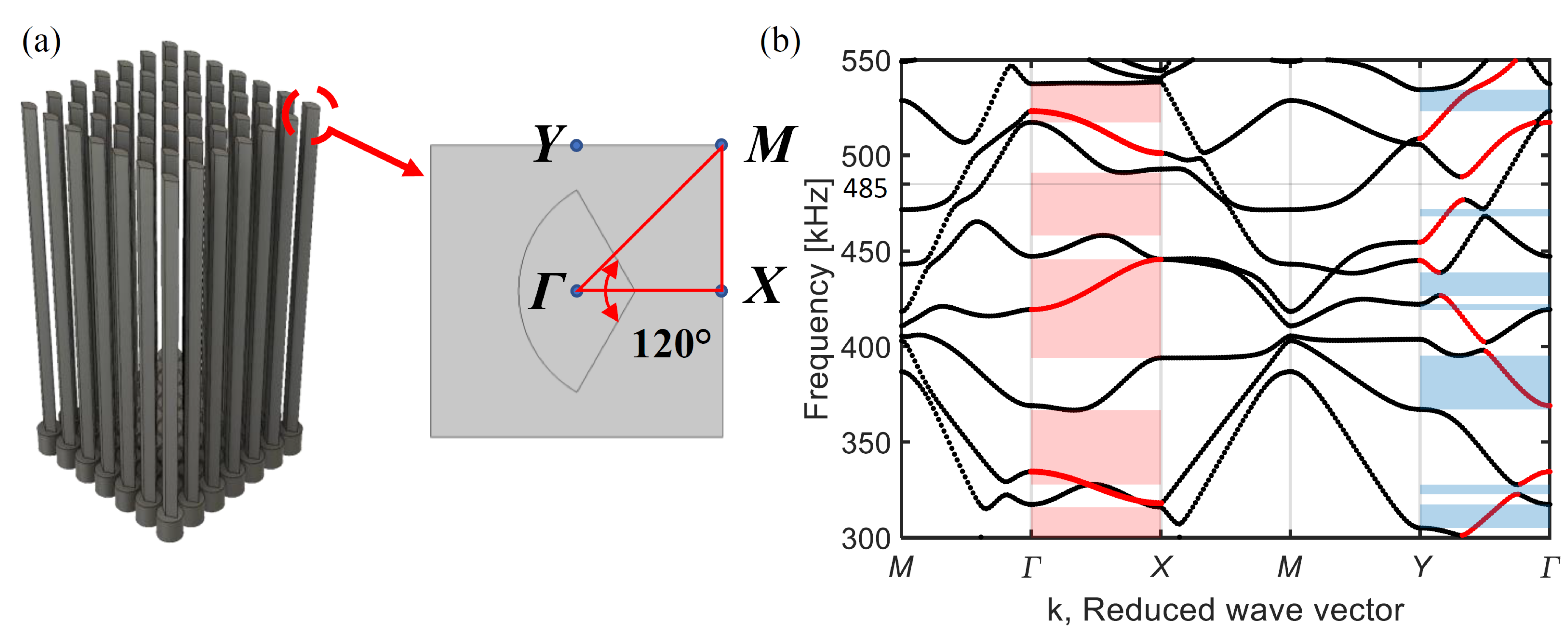}
\caption{Band structure. (a) Phononic crystals with asymmetric scatterers and its unit cell. The irreducible part of the Brillouin zone in $k$-space is inside the red triangle.  (b) Band structure of the PnC with asymmetric rods. Partial band gaps are shaded by different colors. Deaf modes for the $\Gamma-X$ (anti-symmetric modes) and $\Gamma-Y$ (symmetric modes) directions are marked in red. The deaf modes are ignored in shading the non-transmitting regions of frequencies. }
\label{fig3}
\end{figure}

A schematic of the PnC with unit cell and the corresponding band structure are shown in Fig. \ref{fig3}. Due to \emph{\emph{asymmetry of the cross-section of the rods}}, the degeneracies related to rotational symmetry of the unit cell are lifted, giving rise to several partial band gaps shaded by different colors in Fig. \ref{fig3}b. Since the scatterers possess the axis of mirror symmetry ($\Gamma-X$ direction in Fig. \ref{fig3}a), the eigenfuncions are characterized by a definite parity, being symmetric or antisymmetric. Depending on the direction of the incoming wave the eigenmodes with different parity are excited.
The eigenmodes with Bloch vector along $\Gamma-X$ direction can be excited by a plane wave if they are symmetric with respect to the direction of propagation. The antisymmetric modes are not excited and they do not contribute to transmission of sound. If the incoming wave comes from the $\Gamma-Y$ direction, it generates one-side perturbation in symmetric unit cell, which can excite only the antisymmetric modes.  Excitation of the symmetric modes require an input from both sides of the sample. The modes which cannot be excited are so-called ``deaf" modes. Dispersion curves of these deaf modes are highlighted in red in Fig. \ref{fig3}b. Since deaf mode does not contribute to transmission, the region of frequencies where only a deaf mode exists is forbidden for sound propagation. In Fig. \ref{fig3}b the forbidden zones for $\Gamma-X$ and $\Gamma-Y$ directions are shaded by different colors. Strong suppression of sound transmission is expected within these shaded regions.

Even in the case of more symmetric rods with square cross-section, different crystallographic directions are not equivalent and the band gaps along these directions do not completely overlap. This property was explored in Ref. [\onlinecite{Gof}] for demonstration of the theoretical possibility of bandgap tuning. Higher asymmetry of the unit cell gives more options for bandgap tuning. Here, rotation of the rods by a finite angle transforms a transmitting device to an insulating medium. The angle of rotation depends on the direction of the incident acoustic beam. \\

\begin{figure}
\includegraphics[width=\linewidth]{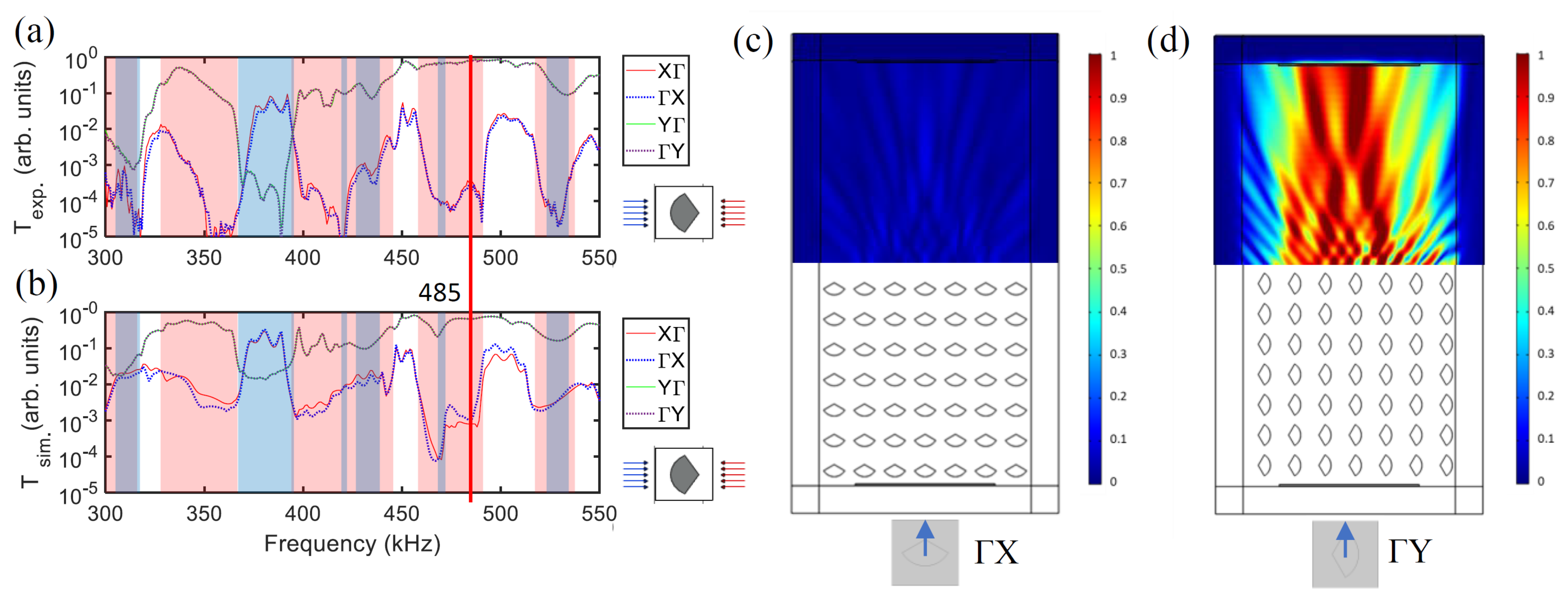}
\caption{Acoustic transmission through the PnC with asymmetric scatterers. (a) Experimental transmission spectra. Vertical red line corresponds to the frequency of 485 kHz. Shaded regions indicate positions of the band gaps shown in Fig. \ref{fig3}. (b) Numerically simulated spectra. Note a good agreement with the experiment. (c), (d) Numerically calculated distribution of acoustic intensity in water for the wave transmitted through the PnC for two mutually perpendicular orientations of the scatterers. The frequency of the incoming wave is $\frac{\omega}{2\pi} =485$ kHz. Along $\Gamma X$ direction the transmission is negligible since 485 kHz fits the bandgap. Unlike this, for $\Gamma Y$ direction the transmission is high. The intensities of the incoming waves (not shown) are equal for both directions. }
\label{fig4}
\end{figure}

Transmission spectra obtained along crystallographic directions $\Gamma X$ and $\Gamma Y$ are shown in Fig. \ref{fig4}. If the direction of the incident beam is fixed, reconfiguration $\Gamma X \leftrightarrow  \Gamma Y$ is obtained by rotating the rods (clockwise or counterclockwise) by $90^{\circ}$. Since the unit cell in Fig. \ref{fig3}a possesses a horizontal axis of mirror symmetry (direction along $\Gamma X$ in $k$-space), the directions $\Gamma Y$ and $Y \Gamma $ are physically identical. Due to mirror or pair symmetry ($P$ symmetry) the transmission spectra along  $\Gamma Y$ and $Y \Gamma $ are indistinguishable. Indeed, experimental spectra in Figure \ref{fig4}a corresponding to these symmetrical directions coincide with high precision. Of course, the numerically simulated spectra in Figure \ref{fig4}b are exactly identical. \\

Mirror symmetry is broken along $\Gamma X$ direction that leads to a noticable difference in the transmission spectra shown in Figure \ref{fig4}a,b by red and blue lines. There are two sources for the difference: asymmetry and nonreciprocity. Asymmetry in transmission appears even for ideal (inviscid) fluid where pressure satisfies Rayleigh reciprocity theorem, $p_A(r_B)=p_B(r_A)$. Here, $A$ and $B$ are two arbitrary points in space where a signal is emitted/received. While the scalar field of pressures is reciprocal, the corresponding vector field of velocities defined through gradient of pressure, ${\bf v(r)} \sim \nabla p(\bf r)$, is not. Acoustic intensity, being a product $I({\bf r}) = p({\bf r}) v({\bf r})$, also does not possess reciprocal symmetry. Since dynamics of an inviscid fluid is reversible, the difference in intensities $I_A({\bf r}_B) - I_B({\bf r}_A)$ associated with broken $P$ symmetry is attributed to asymmetry of scattering but not to nonreciprocity \cite{Maz}. \\

\medskip
A truly nonreciprocal contribution to transmission is due to finite viscosity of the background fluid. If $P$ symmetry is broken, the asymmetry in distribution of velocities is accompanied by different level of dissipated power, which is defined by the gradients of velocities $\left(\partial v_i/\partial x_k\right)$ \cite{LL}.
\begin{equation}
\label{diss}
\dot{Q}= - \int \left[\frac{\eta}{2}\left(\frac{\partial v_i}{\partial x_k} + \frac{\partial v_k}{\partial x_i} -\frac{2}{3} \delta_{ik} \frac{\partial v_l}{\partial x_l} \right)^2 + \xi \left( \nabla \cdot {\bf v} \right)^2\right] dV.
\end{equation}
Viscous fluid dynamics is irreversible, i.e., time-reversal symmetry ($T$ symmetry) is broken. While it was known that broken $PT$ symmetry leads in general to nonreciprocal transmission, viscous dissipation was not considered as a source of nonreciprocity \cite{Maz,Fle} since pressure remains reciprocal even in a viscous fluid. The nonreciprocity induced by viscous losses has been recently demonstrated in propagation of acoustic \cite{Walk,Heo}, elastic \cite{Tan}, and light \cite{Huang} waves. In these studies nonreciprocity is associated with difference in transmission and absorption for propagation of wave along the direction of broken $P$ symmetry.

Usually dissipation-induced nonreciprocity in acoustic transmission is small \cite{Walk,Heo} because time-reversal symmetry is only weakly broken by relatively low dynamic viscosity of water, $\eta \approx 10^{-3}$ N s/m$^2$.
Nonreciprocity can be enhanced by increasing surface roughness of the scatterers. Viscous dissipation mainly occurs within a narrow boundary layer of thickness $\delta = \sqrt{2\eta/\rho \omega}$ formed around each solid rod. Any unevenness along an otherwise flat surface perturbs the velocity field towards stronger gradients, increasing viscous losses Eq. (\ref{diss}). To enhance dissipation-induced nonreciprocity, sandpaper was superglued to the flat sides of each rod. With sandpaperthe mean amplitude of surface roughness becomes about 140 $\mu$m, comparable to $\delta$. At the same time, this roughness is much less than sound wavelength, i.e., it does not materially affect the scattering cross-section of the rods. Thus, the presence of sandpaper does not change much the asymmetric part of the transmission, but enhances the nonreciprocal contribution to the transmission spectra in Fig. \ref{fig4} a,b.

Nonreciprocity is a size-effect, i.e., it disappears for infinite phononic crystal. It is obvious that in an infinite sample viscosity attenuates sound amplitude to zero, independently of the direction of propagation. Moreover, asymmetry of the unit cell and viscous background do not lead to nonreciprocity in the dispersion and attenuation of sound, keeping the relation $\omega(-{\bf k}) = \omega({\bf k})$ valid for both, real and imaginary part of $\omega$, Ref. [\onlinecite{Heo}].  Nonreciprocity may be manifested through the dispersion relation $\omega(-{\bf k}) \neq \omega({\bf k})$ if time-reversal symmetry in a periodic dissipationless structure is broken by temporal modulation of elastic parameters \cite{Norr}.

The forbidden regions marked in the band structure in Fig. \ref{fig3}b are clearly manifested by deep minima in the transmission spectra in Fig. \ref{fig4}a,b. The forbidden zones are shaded in the same colors as in Fig. \ref{fig3}b. Narrow gaps for the $\Gamma-Y$ direction are not resolved for this relatively short sample.  The forbidden regions for $\Gamma-X$ and $\Gamma-Y$ directions partially overlap. For the regions of frequencies where gaps do not overlap, the proposed active phononic crystal serves as a valve, switching from an opaque to transparent state by synchronically rotating the rods by $90^{\circ}$. Distribution of intensity in Figure \ref{fig4} (c) and (d) numerically calculated for the frequency of 485 kHz demonstrates increase in the transmission by 3 orders of magnitude when the rods change configuration with respect to the incoming beam.

In passive structures, sharp change in transmission requires a rotation of the entire crystal. Passive phononic structures demonstrated in previous reports can provide either a conducting state or an insulating state \cite{Gof}. Active modification of the intrinsic materials properties has shown to have a limited influence on the tunability of transmission properties \cite{Zhang}. In contrast, the mechanically reconfigurable PnC serves as a robust acoustic valve, enabling fast switching between insulating and conducting regimes and may also continuously regulate the intensity of the transmitted beam.

\subsection{Tunable gradient index lens}
Focusing of sound can be achieved in the same standard way as focusing of light. A divergent beam passing through a convex phononic crystal with effective index of refraction higher than that of the ambient medium becomes a convergent beam \cite{Cerv}. Here, the phononic crystal operates in the regime of homogenization, i.e., serving as a homogeneous medium with prescribed elastic properties \cite{Kro}. The elastic properties may vary with coordinates by smooth variation of, e.g., the filling fraction. The corresponding effective medium becomes a gradient index phononic crystal where a propagating acoustic wave undergoes continuous refraction caused by coordinate-dependent index.

\medskip

A coordinate-dependent index allows focusing of sound by a flat 2D \cite{Clim} or 3D lens \cite{Sanchis}. Flat lenses have found numerous applications in subwavelength imaging. Phononic crystal bandstructures often have passing bands with negative group velocity above the homegenization limit. Incident sound of frequency within one of these passing bands undergos negative refraction. Negative index metamaterials have served as the ``perfect" lens \cite{Pendry}. The effect of negative refraction has been actively explored for design and fabrication of flat acoustic lenses exhibiting sub-diffraction resolution due to their ability to pick up evanescent modes that break the diffraction limit \cite{Negative}. A modern review on acoustic imaging and focusing can be found in Ref. [\onlinecite{Ma}].

\medskip

Special efforts have been made to design tunable focal length lenses. An original dynamic scheme of focusing to a prescribed point without changing configuration and properties of scatterers was proposed in Ref. [\onlinecite{Yin}]. The acoustic lens consisted of a slab with randomly distributed scatterers. The emitted signal was concentrated after multiple scattering events to some point on the other side of the slab. By manipulating the temporal form of the emitted signal, the sequence of the scattering events was sufficiently controlled to shift the desired focal point. Mechanical control has also been utilized for tuning focal lengths by reorientation of either the individual scatterers or the entire crystal as a unit. Refraction control and tunable focusing of sound was numerically simulated for a 2D PnC with scatterers containing square cross-sections \cite{Feng}. The authors of Ref. [\onlinecite{Ku}] explored the impact of rotational symmetry of the lattice and asymmetry of the inclusions on the tunability of phononic crystal bandstructure.

\medskip

Whereas signal modulation and reorientation of an entire crystal are more passive in nature as the ``lens" itself is unmodified, active tuning schemes leverage changing the properties of the crystal itself for operation. Active tunable focusing using the temperature sensitivity of a hydrogel embedded into a PnC was explored in \cite{Walker1}, where the focal plane was shifted by temperature variations. Here, mechanical manipulation of the individual scatterers in the lattice constitutes active tuning as the properties of the crystal itself are independent of the emission source properties or the spatial relation between the emission source and the crystal as a whole. The asymmetric scatterers in the PnC shown in Figure \ref{fig1} are independently oriented along a desired direction. This provides a capability to construct an indefinite number of scattering environments for an incident sound wave. The combination of asymmetric inclusions with independent control means the local refractive index can be tailored to not only produce lensing, but tuning of the focal spot size and focal length based on the relative orientation from scatterer to scatterer.

\medskip

Equifrequency contours (EFC), i.e. curves $\omega(k_x,k_y)=const$, serve as useful tools to interpret refraction in periodic structures. Strong asymmetry of the unit cell scatterer results in EFC topology and band structure that vary greatly based on orientation of the scatterers with respect to the crystallographic axes and the direction of the incident beam. Here, we analyze refraction for two mutually perpendicular orientations. Intermediate orientations that exhibit properties including multi-refraction/reflection, negative refraction, and backscattering amongst others are left for separate publications.

\medskip

\begin{figure}
\includegraphics[width= \linewidth]{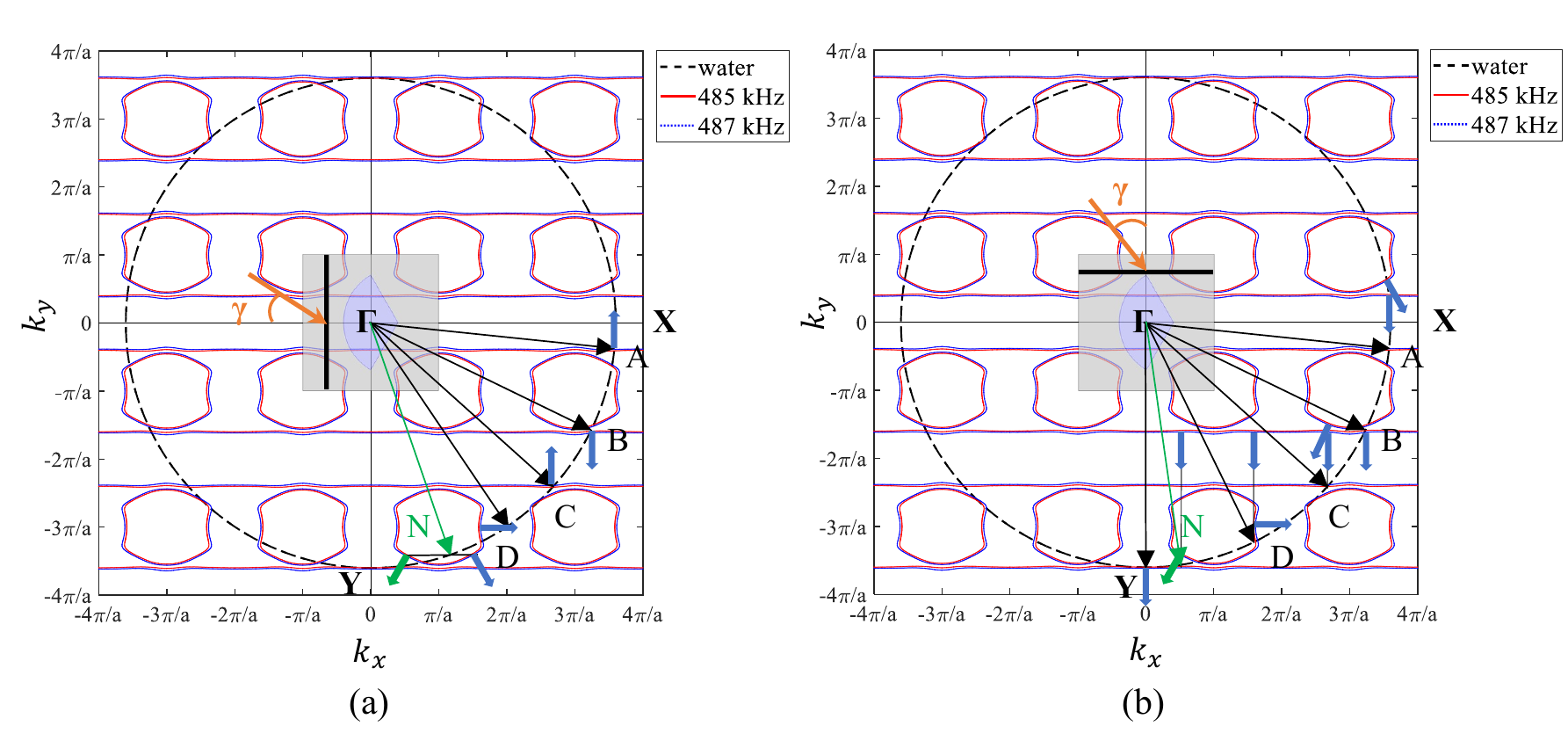}
\caption{Equifrequency contours (EFC) $\omega(k_x,k_y) = 485$ kHz (red lines) and $\omega(k_x,k_y) = 487$ kHz (blue lines) for the phononic crystal with anisotropic scatterers. Refraction of incident plane wave marked by an orange arrow is shown for two orthogonal orientations of the water-crystal interface. Inset at the center is the unit cell. Position of the water-crystal interface is shown by solid black vertical (a) and horizontal (b) lines. Large dashed-line circle is the EFC of water at 485 kHz. Wave vectors of the incident wave at different angles of incidence $\gamma$ are indicated by black arrows. The corresponding group velocities of the refracted waves directed perpendicular to the EFCs are given by bold blue arrows. The geometry of negatively refracted wave is shown by green arrows.  }
\label{fig5}
\end{figure}

In a square lattice, rotation of anisotropic scatterers by $\theta = 90^{\circ}$ does not change the shape of the EFCs. However, diffraction of an  incoming wave at a set of anisotropic scatterers depends on the direction of incidence and orientation of the water-crystal interface. In Fig. \ref{fig5}a the interface is vertical and in Fig. \ref{fig5}b it is horizontal.  In Figure \ref{fig5} the EFC for water at $f=485$ kHz is represented by a circle (black dashed line) of radius $k_0 = 2 \pi f/c_0$, where $c_0$ is speed of sound in water. The EFCs for the PnC are represented by  solid red curves for $f =485$ kHz in the extended-zone-scheme. The contours for slightly higher frequency $f=487$ kHz are shown by dotted blue curves. The EFCs in Fig. \ref{fig5} consist of an unclosed very flat line and a closed curve with concave and convex regions. Note that the flat lines are not closed since there is a wide band gap along $\Gamma X$ direction in Figure \ref{fig3}b.
Since the size of the contours expands with frequency, the direction of group velocity ${\bf v_g} = \nabla \omega({k_x,k_y})$ is outward, perpendicular to the contour. At the crystal-water interface, the direction of the wave vector changes, provided conservation of its tangential to the interface component $k_{\|}$. If the angle of incidence $\gamma$ is counted from the normal to the interface, the condition
\begin{equation}
\label{1}
k_0 \sin \gamma = k_{\|} + \frac{2\pi}{a} n, \,\,\, n=0,\pm1,\pm2,\dots
\end{equation}
defines possible directions of the quasi-wave vector ${\bf k} = (k_x,k_y)$ in the crystal.

\medskip

We assume that in the case of vertical interface $x=0$, the incoming wave marked by orange arrow in Figure \ref{fig5}a enters the PnC from the region $x<0$ with wave vector ${\bf k}_0=(k_0 \cos \gamma >0, -k_0 \sin \gamma<0$). Solutions of Eq. (\ref{1}) with negative (positive) $x$-projection of the group velocity correspond to refracted (reflected) wave. The flat line is along axis $x$ that rules out coupling of external sound to a Bloch wave in the PnC, if the angle of incidence $\gamma$ is within the angle $\angle X \Gamma A$. The same is true for $\gamma$ lying within $\angle B \Gamma C$. For these values of $\gamma$ Eq. (\ref{1}) does not have solution for $k_{\|} = k_y$, i.e., the incoming waves do not penetrate inside the PnC, see, e.g., Figure \ref{fig4}c. If $\gamma \geq \angle X \Gamma A$, then coupling becomes possible.

\medskip

At the point $A$ where the circle intersects with the EFC of the crystal, the phase velocities are equal, i.e., the incoming wave propagating in water along the direction $\overrightarrow{\Gamma A}$ does not suffer phase velocity refraction at the interface. While the wave vector and phase velocity of the refracted wave are parallel to those of the incident wave, the group velocity (shown by blue bold arrow) is perpendicular to the flat line, i.e., practically parallel (or anti-parallel) to axis $y$. Flat part of EFC may be a source of wave collimation, if the same refracted wave corresponds to a continuum of states of incident wave \cite{Christ}. It is not the case for the flat regions shown in Fig. \ref{fig5}a where only three incoming states with $\gamma = \angle O \Gamma A$, $\gamma = \angle O \Gamma B$, and  $\gamma = \angle O \Gamma C$ are directly coupled to the states belonging to the flat regions. Although a normally incident wave is reflected due to partial gap along $\Gamma X$ direction in Fogure \ref{fig3}b, there is an eigenmode in the PnC propagating normally to the interface. This occurs for the incoming wave with $\gamma = \angle X \Gamma D$, which propagates in the PnC with group velocity parallel to the axis $x$. If Eq. (\ref{1}) has more than one solution, there are several refracted waves excited in the crystal. One of possible configurations of multi-refraction is demonstrated by the wave vector $\overrightarrow{\Gamma N}$ shown by green arrow. This incoming wave excites two refracted waves, one of which exhibits positive (blue arrow) and another one -- negative (green arrow) refraction.

\medskip

Although the EFCs remain the same, scattering at the horizontal interface $y = 0$ exhibits very different pattern. In Figure \ref{fig5}b the incoming wave comes from the region $y>0$ with a wave vector ${\bf k}_0= (k_0\sin \gamma >0,  -k_0 \cos \gamma <0)$. The angle $\gamma$ is counted from axis $y$. Since the flat EFC is parallel to axis $x$, Eq. (\ref{1}), where $k_{\|}=k_x$, always has at least one solution for the refracted wave with negative $y$-projection of the group velocity, thus providing coupling with a Bloch wave in the crystal. All the refracted waves originating from the flat line propagate along axis $y$, a signature of strong collimation. The collimated beam propagating perpendicular to the crystal-water interface exists not only at normal but at any oblique incidence. For a wide range of angles $\gamma$ there are two refracted waves in the crystal where one always propagates along the normal to interface, see the incoming waves with wave vectors $\overrightarrow {\Gamma A}, \overrightarrow {\Gamma C}, \overrightarrow {\Gamma N}, \overrightarrow {\Gamma D}$. The incoming wave $\overrightarrow {\Gamma D}$ is strongly refracted  and is coupled to two Bloch waves propagating in the crystal in two orthogonal directions -- parallel and perpendicular to the interface. Negative refraction occurs for the wave incoming along $\overrightarrow {\Gamma N}$.
\medskip

The wave vectors of all incoming waves have the same length $\omega/c_0$. The length of the wave vector of the refracted wave depends on the direction of propagation, as can be seen from  Figure \ref{fig5}. This means that the index of refraction for phase velocity $n_{ph}$  depends on the angle of incidence $\gamma$. It also depends on frequency. This is a signature of metamaterial. Since the wave vector in water exceeds that in the crystal, the index of refraction   $n_{ph}=\frac{c_0}{c_{cr}}=\frac{k_{cr}}{k_0} <1$, i.e. the phononic crystal turns out to be an acoustically less dense  medium than water. The group velocity index of refraction also strongly depends on $\gamma$. Defined by Snell's law for group velocity
\begin{equation}
\label{2}
\frac{\sin \gamma}{\sin i_r} = n_g,
\end{equation}
where $i_r$ is the angle of refraction of group velocity, it is given by the following formula \cite{Christ2}:
\begin{equation}
\label{3}
n_g = - \frac{k_{\|}}{k_0 \cdot \left( \frac{\partial k_{\bot}}{\partial k_{\|}}\right)} \sqrt{1+ \left( \frac{\partial k_{\bot}}{\partial k_{\|}} \right)^2}.
\end{equation}
Here  $k_{\|}$ and $k_{\bot}$ are the tangential and normal to the interface components of the Bloch vector, and the derivative $ \frac{\partial k_{\bot}}{\partial k_\|{}}$ is calculated from the implicit relation $\omega = \omega(k_{\bot},k_{\|})$.

\medskip

\begin{figure}[t]
\includegraphics[width= \linewidth]{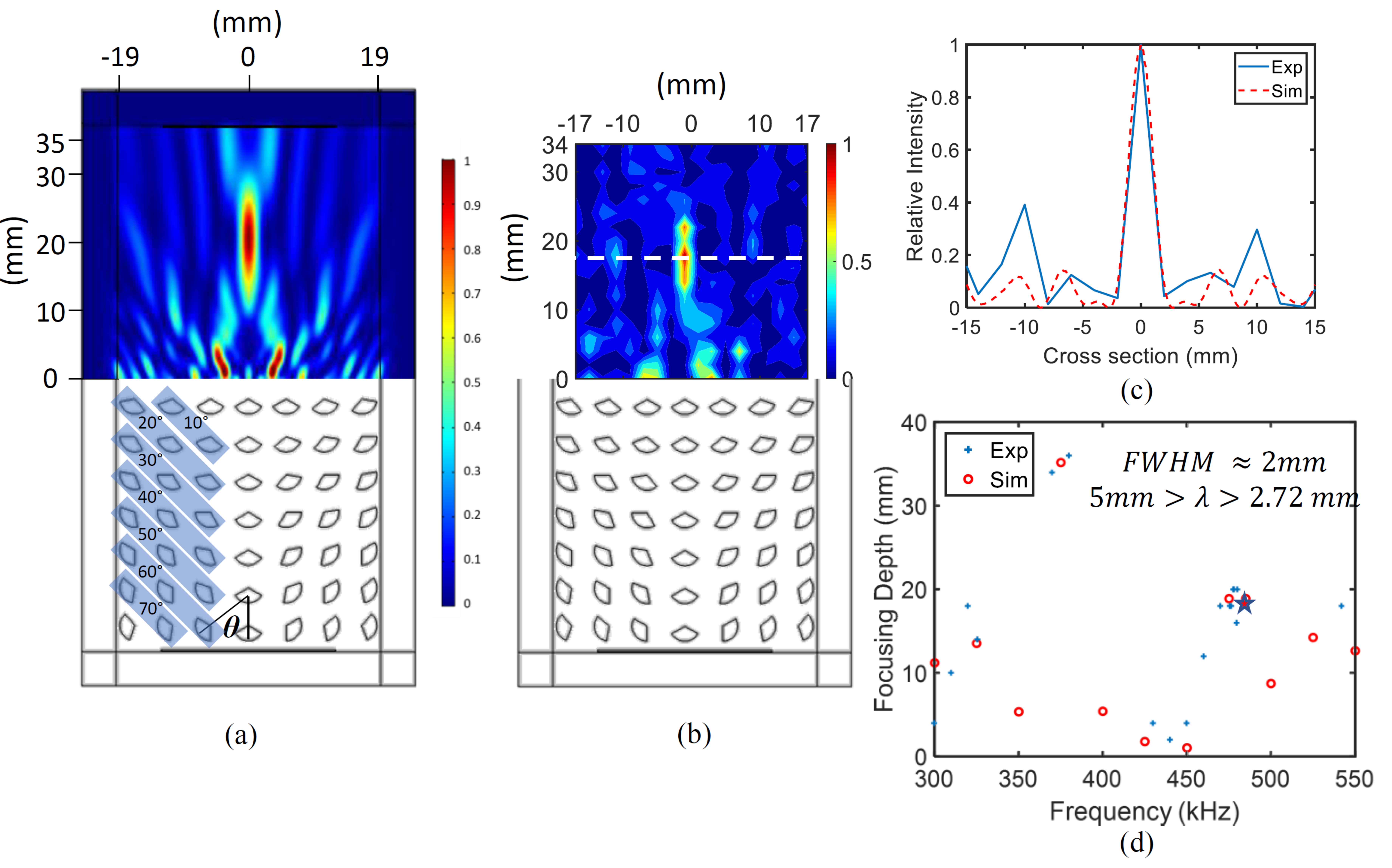}
\caption{Focusing of sound by tunable gradient index lens at 485 kHz. The scatterers are symmetrically distributed with respect to the $y$ axis.  The wave propagates from the bottom to the top. The acoustic intensity map is (a) calculated  numerically and (b) measured experimentally. (c) Decay of intensity along the white dashed line passing through the focal point. In the intensity map the depth and the width of the focal point are 18 mm ($3 \lambda$) and 2.2 mm ($0.72 \lambda$) at frequency 485 kHz. (d) The map showing variation of the depth of the focal point with frequency. The star corresponds to 485 kHz.  }
\label{fig6}
\end{figure}
Equation of the flat EFC in Figure \ref{fig5} is $k_y =const$. For vertical and horizontal orientation of the interface, $k_y = k_{\|}$ and $k_y = k_{\bot}$ respectively. Hence, for the flat contour in Figure \ref{fig5}a $\frac{\partial k_{\bot}}{\partial k_{\|}}= \infty$ and Eq. (\ref{3}) gives $n_g = -k_y/k_0$. For the points $A$, $B$, and $C$ where $k_y = -k_0$. Snell's law (\ref{2}) gives the group angle of refraction $i_r$ to be $90^{\circ}$. This coincides with the geometrical solutions of Eq. (\ref{1}) obtained in Figure \ref{fig5}a. For the flat contour in Figure \ref{fig5}b the derivative $\frac{\partial k_{\bot}}{\partial k_{\|}}= 0$ that means $n_g= \infty$. This results in strong collimation because the angle of refraction $i_r = 0$ for any value of $\gamma$. Thus, with respect to group velocity the phononic crystal in Figure \ref{fig5} behaves like extremally dense acoustic medium. Note that the same phononic crystal refracts phase velocity as less dense than water medium.

\medskip

The rotational degree of freedom of anisotropic scatterers allows one to continuously change the band structure, EFCs, and the refractive properties of the PnC. Orientation of the scatterers can be characterized by angle $\theta$, as shown in Figure \ref{fig6}a.
Since the index of refraction depends on orientation of anisotropic scatterers, the angle $\theta$ serves as a variable parameter in gradient index acoustic lens. Prior gradient index metamaterials leveraged filling fraction as the dependent parameter and required fabrication of scatterers of varying size and/or shape\cite{Clim}. In contrast, the use of uniform anisotropic scatterers and angle dependence reduces fabrication complexity, a potential advantage.

\medskip
We propose the gradient index focusing lens shown in Figure \ref{fig6}. In each row, the scatterers are symmetrically distributed with respect to the central vertical column, where all the rods are equally oriented with $\theta = 0$. A bandgap along this direction prevents direct propagation of sound through the lens, thus reducing the intensity at the lens axis behind the last row of scatterers.  The rods in the first three rows are equally oriented with $\theta = 60^{\circ}$, in the next two, $\theta = 40^{\circ}$, and in the last two rows $\theta = 20^{\circ}$. A monochromatic plane wave is incident normally to the first row from a source at the bottom of the lens and is subsequently focused passing through the gradient index structure. Numerically calculated distribution of sound intensity in water behind the lens for frequency 485 kHz is shown in Fig. \ref{fig6}a. A focal point with 3-5 times higher sound intensity than the ambient is well manifested. The experimental pattern in Fig. \ref{fig6}b is in agreement with the numerical results. The difference in the position of the focal point in the experiment and in the numerical simulations is less than $4\%$.

\begin{figure}[t]
\includegraphics[width= \linewidth]{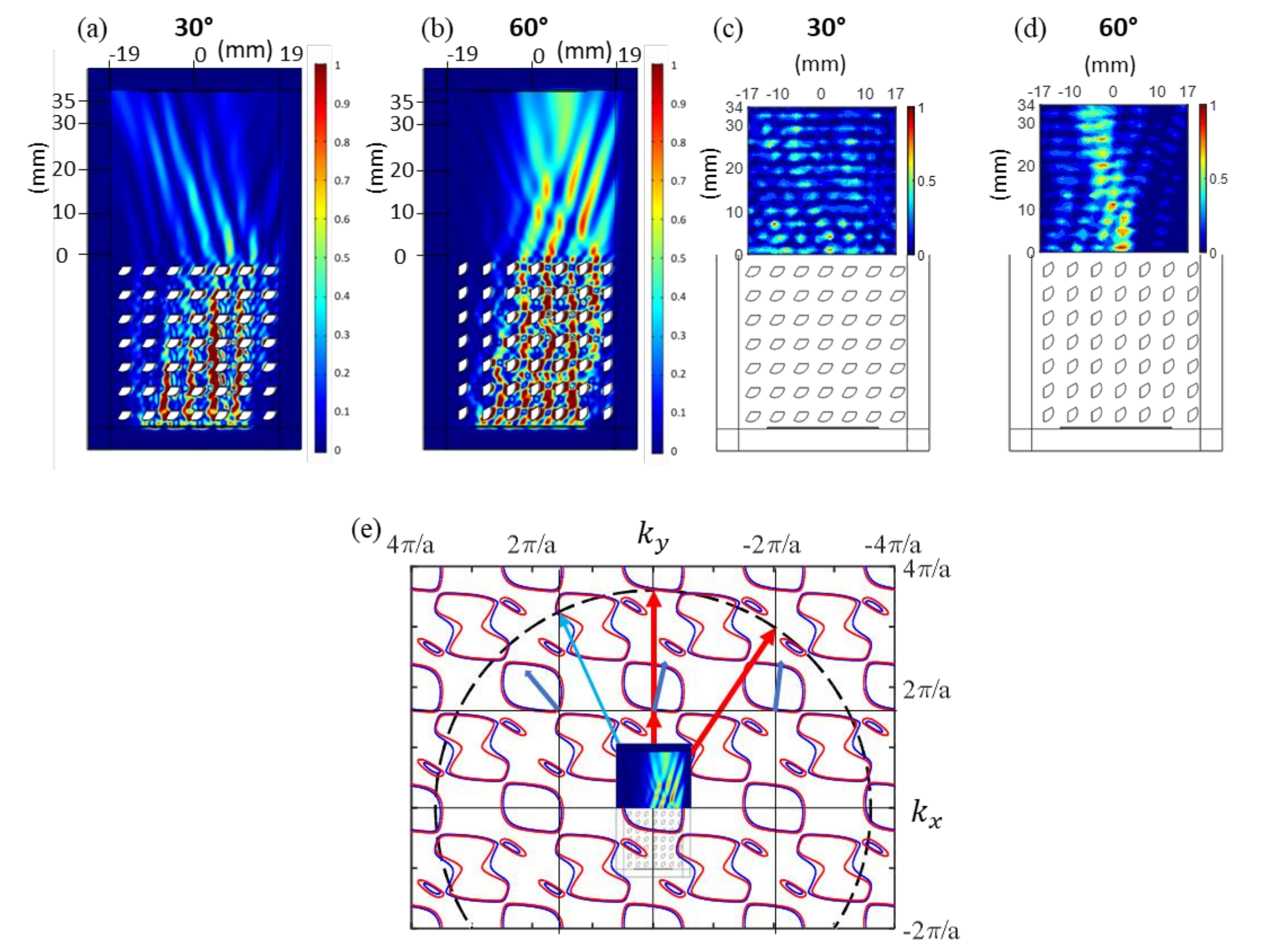}
\caption{Multi-refraction and beam steering for the phononic crystals with scatterers oriented at angles $\theta = 30^{\circ}$ and $\theta = 60^{\circ}$. The incoming plane wave of frequency 485 kHz incidents normally at the lower edge of the crystal. The acoustic intensity maps (a) and (b) show the numerical results and the maps (c) and (d) are obtained experimentally.   (e) EFCs for the orientation $\theta = 60^{\circ}$ and geometrical solutions of Eq. (\ref{1}) with $n=0,\pm1$. The map (b) is placed at the center showing the directions of the refracted beams. }
\label{fig7}
\end{figure}

\medskip

The focal length of the PnC measured from the outer edge of the crystal is 23.5 mm ($\approx 8\lambda$) including the gap, 5.5 mm ($1.8\lambda$), between the hydrophone (detector) and the PnC. According to the Rayleigh criterion of the resolution of the acoustic imaging, the lateral resolution is 2.2 mm ($0.72 \lambda$).
 As shown in Figure \ref{fig6}d, the depth of the focal point varies with frequency due to the incident wave dispersion while traversing through the crystal. Each focusing point is measured by sweeping the frequency at a 1 kHz interval. The tunability is about 35 mm between 300 kHz and 550 kHz having a similar lateral FWHM of approximately 2 mm. Numerical and experimental results for the position and width of the focal point in Fig. \ref{fig6}d are in close agreement with each other.

 \subsection{Beam steering}

In addition to functioning as a transmitter or insulator (Figure (\ref{fig4}c,d)) or acoustic lens (Figure \ref{fig6}), the multi-functionality of the present PnC with anisotropic scatterers extends to beam steering as well. When the scatterers are synchronically rotated  such that angle $\theta$ is uniform, each orientation possess unique dispersion that results in effectively a continuum of alternative media between the limiting cases of transparent and opaque metamaterials shown in Figure \ref{fig4}c,d. Two examples of beam steering for orientations with $\theta = 30^{\circ}$ and $\theta = 60^{\circ}$ are given in Figure \ref{fig7}a,b (numerical simulations) and in Figure \ref{fig7}c,d (experiment).

\medskip

For each orientation, the distribution of transmitted intensity is represented by different patterns. The transmitted wave is split into three beams with different intensities. Such splitting represents the effect of multi-refraction in periodic elastic medium previously reported in Ref. [\onlinecite{Christ}]. Note, that multi-refraction originates from the solutions of Eq. (\ref{1}) corresponding to different value of integer $n$. The directions of the transmitted beams shown in Figure \ref{fig7}c are obtained from the geometrical solutions of Eq. (\ref{1}). There is a reasonable agreement between the intensity maps obtained numerically and experimentally. The directions of the main transmitted beams in Figures \ref{fig7}a and c and  in Figures \ref{fig7}b and d coincide well, however there is some discrepancy in the level of intensity amplitudes.

\medskip

The geometrical solutions of Eq. (\ref{1}) are presented in Figure \ref{fig7}e. The wave vector of the wave incident normally from water is the red vertical arrow touching the water circle. The wave vector of the refracted wave, obtained from Eq. (\ref{1}) for $n=0$, is the shorter red arrow in the same direction (since the component tangential to the surface is zero) touching the nearest EFC. The group velocity of the Bloch wave in the crystal shown by a bold blue arrow is slightly tilted from  the vertical direction. The Bloch wave with these group velocity and wave vector suffers inverse refraction at the upper edge of the crystal. The wave transmitted to free water has the same wave vector as the incident wave and propagates normally to the interface.

\medskip

There are also solutions of Eq. (\ref{1}) corresponding to the 1st-order refracted modes with $n =\pm 1$. In Figure \ref{fig7}e these solutions belong to the EFCs shifted to the left and to the right from the central contour (partially covered by image of the PnC sample) by one inverse lattice period. Directions of the group velocities of the corresponding refracted Bloch waves are shown by bold blue arrows. The Bloch waves at the right and central contours have equal group velocities. Refraction at the upper edge of the PnC occurs with conservation of the tangential (horizontal) component of the wave vector. This requirement gives two vectors for the transmitted waves lying at the water circle and shown by the red and blue arrows in Figure \ref{fig7}e. The corresponding waves propagate in free water under different angles with respect to the incident wave. The directions of the red and blue arrows coincide well with the directions of the main beams in Figures \ref{fig7}b and d obtained numerically and experimentally.  The EFCs of the PnC with $\theta = 30^{\circ}$ are topologically equivalents to the contours in Figure \ref{fig7}c. We omit here the geometrical procedure that defines the directions of the transmitted beams for $\theta = 30^{\circ}$  since it is very similar to that shown in Figure \ref{fig7}e. Due to asymmetry of the scatterers, the intensities of the tilted beams in Figure \ref{fig7} are not equal, allowing the highest intensity beam to be steered smoothly with orientation.

\medskip

\begin{figure}
\includegraphics[width= \linewidth]{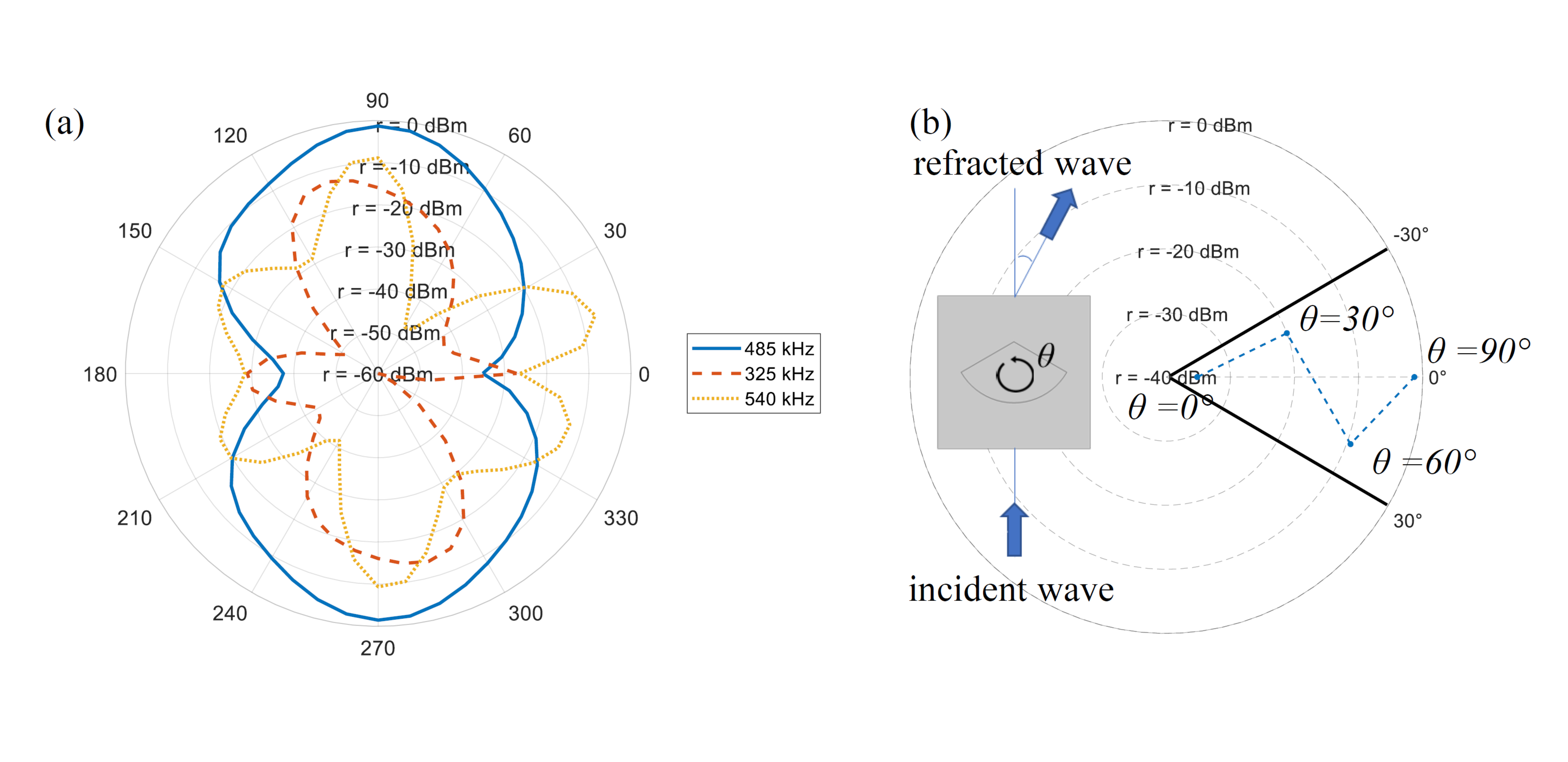}
\caption{Experimental results for beam steering at frequencies 325, 485, and 540 kHz. (a) Maximum of intensity in the transmitted wave vs angle of orientation $\theta$. (b) Polar diagram for intensity versus steering angles at 485 kHz. The radius of the polar plots in (b) denotes the level of the intensity with the dBm scale measured by the experiment. Inset in (b) shows the angle of refraction through the unit cell and the rotation of the unit cell. }
\label{fig8}
\end{figure}

 The experimental demonstration of beam steering was conducted at frequencies 325, 485, and 540 kHz for the whole range of angles $\theta$. The polar diagram in Figure \ref{fig8}a shows how the intensity of the transmitted beam changes with angle of orientation $\theta$. At frequency 485 kHz, maximum intensity is reached for $\theta = 90^{\circ}$ and minimum -- for $\theta = 0^{\circ}$, in agreement with the transmission pattern in Figure \ref{fig4}c,d. At frequencies 325 and 540 kHz multiple minima and maxima of intensity are observed  for different orientations. Note, since the scatterers have axis of mirror symmetry,  the diagrams exhibit mirror symmetry with respect to the horizontal axis and there is no symmetry with respect  to the vertical axis.

 \medskip

The steering effect of rotating synchronized asymmetric scatterers can be seen in the polar plot in Figure \ref{fig8}b.  For the normally incident wave with frequency 485 kHz the direction of the main beam follows the rotation angle for $\theta<40^{\circ}$. However, for larger angles the steering direction switches, i.e., counterclockwise rotation of the scatterers leads to clockwise tilt of the main beam.   For $\theta<90^{\circ}$ the transmission reaches its maximum (see Figure \ref{fig4}d) and the transmitted beam returns to the direction of the incident wave. The steering of sound occurs within the interval of angles from approximately $-30^{\circ}$ to $30^{\circ}$ when the angle of rotation of scatterers is within the range $0<\theta < 90^{\circ}$. Note that the amplitude of transmitted sound strongly depends on steering direction.  Electrically configurable metaslab in Ref. [\onlinecite{Popa}] provides beam steering within the same interval of angles.

\section{Conclusions}
This study demonstrates an active phononic metamaterial device with multifunctional characteristics. Due to the presence of rotational degree of freedom and asymmetry of the scatterers, the device can be used as transmitter or insulator of sound, gradient index lens, and beam steerer. Switching from one regime to another is achieved in real time due to electrical motors controlling orientation of each scatterer.
Various direct applications can be proposed in selective filtering of the acoustic power, energy concentration and harvesting, communications and imaging systems, signal processing, and thermal management. Combining the active mechanism and nonreciprocity can provide a system to trap/dump sound energy. This approach can be extended to higher or lower frequency bands by scaling down or up the size of the crystal. Continuous synchronized rotation of scatterers can be considered as adiabatic temporal variation of elastic parameters of PnC. It was shown \cite{Norr} that such slow modulation in 1D periodic structure leads to quantized tilt of PnC bands. The tilted bands lose the parity symmetry, which is a fundamental property of time-independent bandstructure. Rotation of periodically arranged scatterers breaks time-reversal symmetry, leading to a non-Hermitian eigenvalue problem. Rotation by itself provides Doppler-induced scattering sufficient for formation of band gaps without acoustic contrast between the constituents of PnC\cite{Chan}. Manipulation with velocity and direction of rotation allows acoustic cloaking in a homogeneous medium\cite{Wu}.  Thus, the proposed device may demonstrate a different mechanism of nonreciprocity, unrelated to viscosity of the background. As well as the aforementioned viscosity-based mechanism \cite{Walk,Heo}, this mechanism requires scatterers, which break time-reversal symmetry in a periodic elastic medium.

\section{Acknowledgements}
This work is supported by an EFRI grant No. 1741677 from the National Science Foundation. AN and JJ acknowledge support from the Ministry of Science and Technology of China (MOST)  through International Collaboration grant No. 2022YFE0129000 entitled “Cavity Acoustodynamics for Nonreciprocal Wave Propagation".

\end{document}